\documentclass[twocolumn,pre,floatfix,showpacs]{revtex4}

\usepackage{epsfig}
\usepackage{amsmath2000}

\providecommand{\abs}[1]{\lvert#1\rvert}

\begin{document}

\title{Stability of charge inversion, Thomson problem and application 
	to electrophoresis }

\author{Michael Patra}
\author{Marco Patriarca}
\author{Mikko Karttunen}
\affiliation{
Biophysics and Statistical Mechanics Group,
Laboratory for Computational Engineering, 
Helsinki University of Technology, P.\,O. Box 9203, FIN-02015 HUT, Finland}

\begin{abstract}

We analyse charge inversion in colloidal systems at zero temperature using
stability concepts, and connect this to the classical Thomson problem of
arranging electrons on sphere.  We show that for a finite microion charge, the
globally stable, lowest energy state of the complex formed by the colloid and
the oppositely charged microions is always overcharged. This effect disappears
in the continuous limit. Additionally, a layer of at least  twice as many
microions as required for charge neutrality is always locally stable.  In an
applied external electric field the stability of the microion cloud is reduced.
Finally, this approach is applied to a system of two colloids at low but
finite temperature.

\end{abstract}


%

\maketitle

\section{Introduction}

We study analytically and numerically charged colloidal particles in the
presence of electrolyte solutions in the low-temperature limit. This is
equivalent to strong electrostatic coupling at finite
temperature~\cite{moreira:00a}. Already at room temperature a colloidal system
thus is in the low temperature regime, provided the electrolyte consists of
multivalent  ions.  In this limit, a certain number of ions condenses onto the
surface of the colloid. For consistency, we will call the charged colloids
macroions, and the ions of the electrolyte microions. 
For the general macroion problems, this behaviour has recently attracted a lot
of attention especially due to its importance in biological 
systems~\cite{manning:69a,manning:69b,oos:70,gelbart:00a,shklovskii:99a,park:99a,grosberg:02a}.

Macroion
complexes exhibit numerous counter-intuitive phenomena. The most pronounced one
is referred to as overcharging or charge inversion. A certain number of
microions is needed to condense for the entire complex (macroion and condensed
microions) to become charge neutral. Sometimes an excessive number of microions
condenses~\cite{allahyarov:98a,shklovskii:99a,park:99a,linse:99a,mateescu:99a,gurovitch:99a,lozada:99a,nguyen:00a,nguyen:00b,messina:00a,messina:00b,messina:01b,tanaka:01a,grosberg:02a,messina:02a,terao:01a}
and the complex, the ``dressed'' macroion, acquires an effective charge that is
opposite in sign to that of the bare macroion. This phenomenon is beyond the
standard Debye-H\"uckel~\cite{neu:99a,sader:99a,trizac:99a}
and Derjaguin-Landau-Verwey-Overbeek (DLVO)
theories~\cite{evans:99}.

Another phenomenon 
is auto-ionisation~\cite{messina:00a,messina:00b,messina:01b}. 
This means that one macroion transfers some of its microions to another
macroion, so that the first one becomes undercharged while at the same time
the second one becomes overcharged. Also analysed in the literature is the
important question of transport in an external electric field. A bare macroion
will move in the direction determined by its own charge. The binding of
microions to it can under certain conditions reverse the direction.

These results have mostly been arrived at by molecular dynamics (MD) and  Monte
Carlo (MC) simulations. Simulations offer the advantage that finite temperature
can be taken into account in a natural way. Purely analytical approaches have
to resort to relatively complicated starting points since mean-field theories
are insufficient~\cite{neu:99a,sader:99a,trizac:99a}.  A successful and often
used approach is  the model of a two-dimensional Wigner crystal which becomes
exact at zero temperature and very large number of microions per
macroion~\footnote{At very large number of microions, the curvature of the
macroion surface seen by each microion becomes negligible. Then and only then
the microion layer becomes two-dimensional.}.

At zero temperature --- without any restrictions on the number of
microions --- the problem is, however, directly 
related to the classical ``Thomson problem''
of finding stable configurations  of $N$ mutually repelling electrons on the
surface of a  sphere~\cite{thomson:04a,erber:91a}.  While the original problem
was about the ``Plum Pudding'' model for the atom, where $N$ particles are 
confined inside a homogeneously charged sphere, both problems actually are
identical since the repelling interaction will push all particles inside the
sphere toward its surface. This similarity between overcharging and the Thomson
problem seems to have been largely unnoticed in the macroion literature. 

In this paper we will make use of the Thomson problem to derive rigorous bounds
for the phenomena discussed above. At zero temperature, our results are exact
and hence an improvement on previously known results from Wigner crystal theory.
Compared to the results of simulations, our derivation suffers from our
inability to include finite temperature in an exact way. However, our method
outperforms previous ones in both the ease of the method (both conceptually
and numerically) and in allowing to treat many phenomena in a single consistent
way.

This paper is organised as follows. In Sec.~\ref{secModel} we summarise the
so-called primitive model. This model is used in basically all
studies of macroion complexes. In Secs.~\ref{secGlobal}--\ref{secVergleich}
we analyse the stability of overcharged macroion complexes. We use concepts
from dynamical-systems theory to show that two different stability properties
exist, global stability (Sec.~\ref{secGlobal}) and local stability
(Sec.~\ref{secLocal}). In Sec.~\ref{secExternal} we move on to the question of
a macroion in an applied external field, i.\,e., electrophoresis. We will
discuss the auto-ionisation of
macroions in Sec.~\ref{secAsymetry}. We conclude in Sec.~\ref{secConclusions}.

\section{Model} 
\label{secModel}

We consider a spherical macroion of charge $Q$ and radius $R_{\mathrm{mac}}$, 
surrounded by $N$ spherical microions of charge $q$ and radius
$R_{\mathrm{mic}}$. $Q$ and $q$ are of opposite signs, and in the following
we assume $Q<0$. The macroion is fixed at the origin 
and the $N$ microions are distributed at positions $\vec{r}_i$, 
$i=1,\ldots,N$. The total electrostatic energy $V$
for a particular configuration is then given by
\begin{equation}
        V(\{\vec{r}_i\}) = \frac{q^2}{4\pi\epsilon} \sum_{i<j}^{1\ldots N}
        \frac{1}{|\vec{r}_i-\vec{r}_j|} 
                + \frac{q Q}{4\pi\epsilon} \sum_{i=1}^N
                \frac{1}{|\vec{r}_i|} \;,
        \label{eqVpot}
\end{equation}
where the first sum accounts for the mutual repulsion of the microions, and the
second for the attraction between the macroion and each microion.
The effect of solvent is included through an effective dielectric
constant $\epsilon$.
Short-range pairwise repulsion is taken into account by
hard-core interaction 
\begin{equation}
        V_{\mathrm{hc}}=\sum_{i=1}^N v(
                |\vec{r}_i| - R_{\mathrm{mac}}-R_{\mathrm{mic}} )
        +\sum_{i<j}^{1\ldots N} v(
                |\vec{r}_i-\vec{r}_j| - 2 R_{\mathrm{mic}} )
        \;,
        \label{eqVhc}
\end{equation}
where $v(r) \rightarrow \infty$ for $r<0$ and zero otherwise.
Equations~(\ref{eqVpot}) and~(\ref{eqVhc}) comprise the so-called primitive
model~\cite{grosberg:02a}.

\section{Global stability}
\label{secGlobal}

We approach the problem by using  the
well-known Earnshaw's theorem~\cite{earnshaw:42a} which states that 
there can be no stable state in
a system with only electrostatic interactions present. For
stable configurations to exist, short-range repulsive forces must be present in
addition to the long-range Coulomb ones. For our system, the short-range
forces are due to hard-core interaction [Eq.~(\ref{eqVhc})].
Earnshaw's theorem thus
restricts stable configurations 
to have all microions at a distance $R\equiv
R_{\mathrm{mac}}+R_{\mathrm{mic}}$ away from the centre of the macroion. 

The condition $|\vec{r}_i|=R$ allows us to simplify Eq.~(\ref{eqVpot}) to
\begin{equation}
        V(\{\vec{r}_i\}) = \frac{q^2}{4\pi\epsilon R} \sum_{i<j}
        \frac{1}{\abs{\vec{r}^{\circ}_i-\vec{r}^{\circ}_j}} 
        + \frac{q Q N}{4\pi\epsilon R} 
        \;,
        \label{eqVpot2}
\end{equation}
with the normalised coordinates
$\vec{r}^{\circ}_i \equiv \frac{1}{R} \vec{r}_i$.
Equation~(\ref{eqVpot2}) no longer describes the energy of an arbitrary
arrangement $\{\vec{r}_i\}$ of particles but the energy of \emph{any stable}
arrangement instead. Next, we introduce the function $f(N)$,
\begin{equation}
        f(N) = \sum_{i<j}^{1\ldots N} \frac{1}{\abs{\vec{r}^{\circ}_i
                -\vec{r}^{\circ}_j}} \quad\mathrm{with}\quad 
                \abs{\vec{r}^{\circ}_i}=1 \;,
        \label{eqFdef}
\end{equation}
where we demand that the coordinates $\{\vec{r}^{\circ}_i\}$ 
are the ones for the lowest energy state with $N$ microions around the
macroion. Thus, the coordinates are completely defined
by $N$.

\begin{figure}
\centering
\epsfig{file=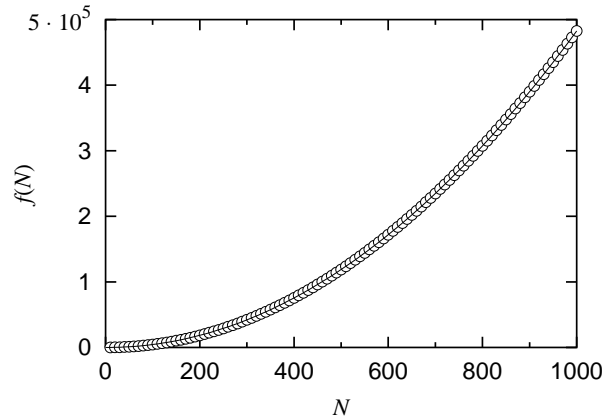,width=3.2in}
\caption{Comparison of Eq.~(\protect\ref{eqfNtheory}) [solid line] with the
result of a numerical computation of Eq.~(\protect\ref{eqFdef}) [open circles].}
\label{figfN}
\end{figure}

At the ground state 
Eq.~(\ref{eqFdef}) becomes minimised. The complete solution
can be computed
numerically very efficiently~\cite{bulatov:96a}. 
Furthermore,
the functional form of $f(N)$ is known to excellent
precision~\cite{erber:91a} to be 
\begin{equation}
        f(N)=\frac{N^2}{2}- c N^{3/2} \quad\mathrm{with}\quad
                        c=0.5510 \;.
        \label{eqfNtheory}
\end{equation}
This formula is easy to understand when one notices that the first term is the 
energy of a continuous layer of charge on a sphere of unit radius while the
second term is the self-energy correction due to discrete microions which can be shown
to be proportional to $N^{3/2}$. 
A comparison of Eq.~(\ref{eqfNtheory}) 
and a numerical solution of the exact 
formula [Eq.~(\ref{eqFdef})] is shown in Fig.~\ref{figfN}. 

By using the condition $\abs{\vec{r}_i}=R$
for all $i$
we have completely
accounted for the hard-core interaction between the macroion and the microions. 
We can neglect the hard-core interaction between microions since they
repel each other as their charges have the same sign.
Collecting results, the potential energy of the lowest energy state for
a macroion surrounded by $N$ microions is given by
\begin{equation}
        V(N) = \frac{q^2}{4\pi\epsilon R} \left[
                \frac{N^2}{2}-c N^{3/2} \right] +
                \frac{q Q N}{4\pi\epsilon R} 
        \;.
        \label{eqPot3}
\end{equation}

Earnshaw's theorem gives a necessary but not sufficient criterion for the
stability of a system. Furthermore, it states that an unstable microion
is immediately pushed to
infinity.
Let us consider a macroion and $N$ microions where we place the microions at
arbitrary positions --- not necessarily on the macroion. Due to Earnshaw's
theorem, $M$ of them will attach to the macroion while $k=N-M$ will escape to
infinity. All $N$ will go to the macroion if it is the state lowest in energy,
i.\,e., if
\begin{equation}
        V(N) < V(M) \quad \forall \quad 0 \le M < N \;.
        \label{eqInV}
\end{equation}
This condition is much stronger than the simple condition $V(N)<0$, since the
latter only prevents \emph{all} microions from escaping simultaneously while
Eq.~(\ref{eqInV}) also prevents \emph{some} from escaping.

Due to Earnshaw's theorem, all stable solutions are enumerated by 
the number of microions, and we simply have to find the number yielding the
lowest energy. Since Eq.~(\ref{eqPot3}) possesses 
only a single extremum for given parameters  $q$, $Q$ and $R$, 
we 
can simply use $d V(N)/d N=0$. That yields
\begin{equation}
        N_{\mathrm{glob}} = \frac{\abs{Q}}{q} + \frac{9}{8} c^2 
                + \frac{9 c^2}{8} \sqrt{1+\frac{16}{9 c^2}
                        \frac{\abs{Q}}{q} } \;,
        \label{eqNstabile}
\end{equation}
where we have used the assumption $Q<0$ introduced above. 

\begin{figure}
\centering
\epsfig{file=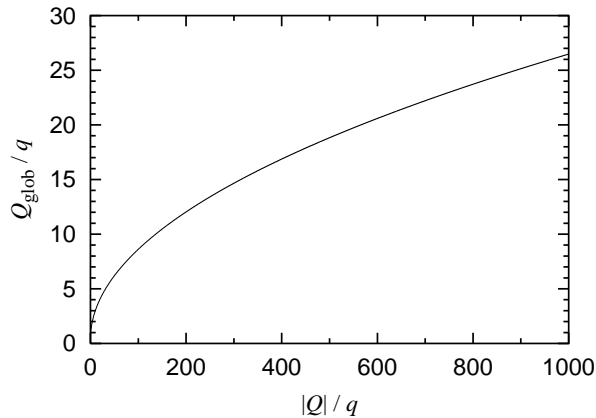,width=3.2in}
\caption{Globally stable overcharging $Q_{\mathrm{glob}}$ as function
of the ratio of the charges of macro- and
microion, computed from Eq.~(\protect\ref{eqNstabile}).}
\label{figN}
\end{figure}

The first term $\abs{Q}/q$ gives the naive result that a complex
consisting of macroion and layer of microions should be charge neutral. The other
two terms give the excess bound microions.
The maximum stable overcharging in terms of charge 
is $Q_{\mathrm{glob}}=q N_{\mathrm{glob}} - \abs{Q}$, 
see Fig.~\ref{figN}.

The stability criterion that we derived applies to an arbitrary
initial placement of microions. For this reason, this kind of stability is
referred to as \emph{global} stability, see any textbook on nonlinear dynamics,
e.\,g. Ref.~\onlinecite{drazin:92}.

\section{Local stability}
\label{secLocal}

In addition to global stability, there exists the concept of \emph{local}
stability. While global stability states that the microions will move to the
macroion independent of their initial positions, local stability means that they
will stay at the macroion if they have initially been placed there.
The system is locally
stable (but not globally) if the system could lower its energy by
transferring one (or more) microions from the macroion to infinity but in doing
so would need to cross an energy barrier. Since we are restricting ourselves to
classical physics at zero temperature, it is impossible to cross such a barrier
and the microions would stay touching the macroion  forever --- if prepared
with this initial condition.

To calculate the condition for the existence of such a barrier we move 
particle $k$ slightly away from the macroion by a distance $\Delta$, keeping
all other microions on the surface of the macroion.  If this move increases the
potential energy, resulting in a restoring force, the system is locally stable.

\begin{figure}[b]
\centering
\epsfig{file=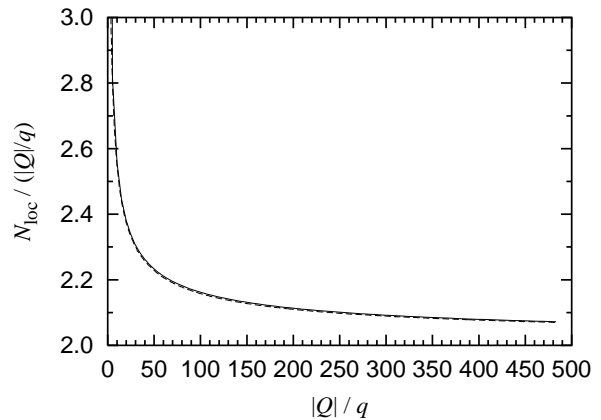,width=3.2in}
\caption{Number of locally stable microions as a function of the ratio of the
charges. The solid line is the analytic upper bound 
Eq.~(\protect\ref{eqUpper}),
while the dashed line is a numerical solution of the exact formula
Eq.~(\ref{eqVpotStrich}).
}
\label{figLocal}
\end{figure}

We label all quantities in the perturbed state by a prime, 
hence $r'_k=R+\Delta$ and $r'_l=r_l=R~\forall~l\ne k$, and we introduce the
abbreviation $d_{kl}=\abs{\vec{r}_k-\vec{r}_l}$. 
Since we need to consider only small
$\Delta$, we can use a series expansion, with the result
\begin{equation}
        \frac{1}{d'_{kl}}=\frac{1}{d_{kl}}-\frac{1}{2 R d_{kl}}
                \Delta \quad\text{and}\quad
        \frac{1}{\abs{\vec{r}'_k}}=\frac{1}{R}-
                \frac{1}{R^2}\Delta \;.
\end{equation}
Inserting this into Eq.~(\ref{eqVpot}) gives 
\begin{equation}
        V' = V - \frac{q}{4\pi\epsilon R}
                \left[ q\sum_{i\ne k}^{1\ldots N} \frac{1}{2 d_{k i}}
                        + \frac{Q}{R} \right] \Delta 
         \equiv V - \frac{h_k}{R} \Delta \;. 
        \label{eqVpotStrich}
\end{equation}
The system is locally stable if and only if the expression in brackets is
negative for all $k$ since then an increase in $\Delta$ will increase the 
potential energy.
Thus, the condition for local stability is $h_k<0 \quad\forall~k$. 
Since the lowest energy arrangement of the particles rarely is completely
symmetric, this yields the necessary but not sufficient condition
$\langle h_k\rangle_k < 0$, where this average is over all possible particles
$k$. Noting that $\sum_k h_k=V(N)$, this gives the
necessary condition for local stability
\begin{equation}
        V(N) < 0 \;.
        \label{vNlocal}
\end{equation}
It should be noted that this simple form for the condition 
is a coincidence, and for other systems $V(N) < 0$ has not necessarily 
relation to local stability.
With the help of
Eq.~(\ref{eqPot3}) this condition can be converted into 
an upper bound for the number of
microions that can be bound locally stable,
\begin{equation}
        N_{\mathrm{loc}} = 2 \frac{\abs{Q}}{q} + 2 c^2 \left[ 1 
        + \sqrt{1+\frac{2\abs{Q}}{c^2
        q}} \right] \;.
        \label{eqUpper}
\end{equation}
In terms of charge this is 
$Q_{\mathrm{loc}} = 
q N_{\mathrm{loc}} - \abs{Q}$.

To check the difference between 
Eq.~(\ref{eqUpper}) and the exact
solution, we have numerically computed  the lowest
energy state as a function of $N$, and from that
determined the largest $h_k$ for each $N$.
The result in Fig.~\ref{figLocal} shows that hardly any difference between
the two values can be seen. This does not come as a surprise since the
differences between $h_k$ for different $k$ are small as
the repelling forces
among the microions try to make all mutual distances as equal as possible.

\section{Summary of stability concepts}
\label{secVergleich}

\begin{figure}[b]
\epsfig{file=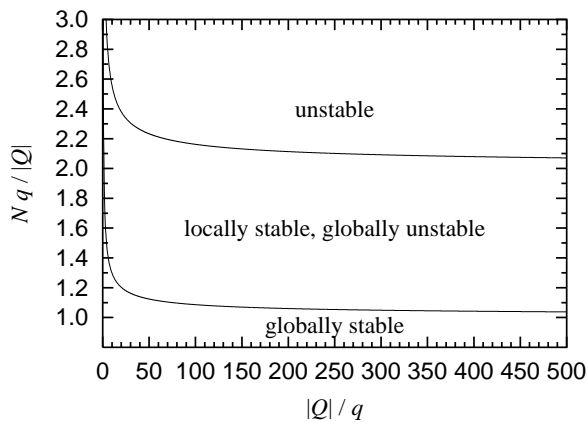,width=3.2in}
\caption{Number of microions that can be bound globally stable or locally stable
to a macroion. $N q /\abs{Q}=1$ is the number of microions
expected from charge neutrality.
}
\label{figComparison}
\end{figure}

For finite $q$, the number of microions that are bound globally stable is 
always
larger than the value $N=\abs{Q}/q$, i.\,e., the macroion is overcharged.
In the continuous limit $q\to 0$ this effect disappears.
In contrast, the number of locally stable bound microions is at least twice 
the amount needed for charge neutrality, and this effect persists even in the
continuous limit. 
Fig.~\ref{figComparison} shows the different regimes as a function of the
charges of macroion and microions.

We now want to put our results into perspective of previous results on
overcharging~\cite{allahyarov:98a,shklovskii:99a,park:99a,linse:99a,%
mateescu:99a,gurovitch:99a,lozada:99a,nguyen:00a,nguyen:00b,messina:00a,%
messina:00b,tanaka:01a,messina:01b,terao:01a,messina:02a,grosberg:02a}.
At finite $T$, there are only few simulations done in the geometry employed in
this paper, e.\,g., Ref.~\onlinecite{terao:01a}.
Most analytical work focuses on $T=0$, both 
due to simplicity and 
allowing one to focus on the influence of correlations. 
We employ  the same approach.  

The main advantage of our approach is the adoption of the exact Thomson model
as source for the potential energy $V(N)$ whereas previous papers used an
energy estimate for $V(N)$ derived from strongly-correlated liquid and Wigner
crystal theory~\cite{shklovskii:99a,messina:00a}. The Wigner crystal theory 
contains a parameter
$\alpha$ whose value can either be determined from simulations as a function of
$N$  --- which trivially leads to a self-consistent result --- or needs to be
fixed with the analytical value for $\alpha$ known for a two-dimensional Wigner
crystal. In the latter case, this results in an error of up to order $10\,\%$
for the computed energies~\cite{messina:00a}. In contrast the error of
Eq.~(\ref{eqfNtheory}) is negligible (less than $10^{-4}$).

We apply methods from nonlinear dynamics, using the concepts of local and global
stability. The existence of these two different stability properties seems to be
unnoticed in macroion literature. For example, the criterion by Messina 
\textit{et al.}
agrees with our global stability criterion up to the differences caused by their
choice for $V(N)$. The concept of local stability, however, is also an important
one as can be seen, for example, in the electrophoresis setup treated in the
following section. 

Finally, for numerical calculations we employ a
minimisation scheme. Only a few different initial conditions are necessary to
make sure that the algorithm does not become stuck in a local minimum.
This is in contrast to MD simulations which
suffer from the slowing down of the dynamics at low temperature.

\section{Macroion in an external electric field}
\label{secExternal}

Let us consider a macroion with $N$ microions in an external field
$\vec{\mathcal{E}}$.
We will restrict ourselves to the case of homogeneous external
field so that the dipole and higher moments of the macroion complex are
irrelevant. 
Typical electrophoresis experiments are done in the
presence of a homogeneous field.

The total force acting on the complex becomes simply
\begin{equation}
        \vec{F} =  ( q N + Q ) \vec{\mathcal{E}} \;.
        \label{eqFtotal}
\end{equation}
The complex will thus move in the same direction as the bare macroion if the
macroion is undercharged, it will move in the opposite direction if it is
overcharged, and it will remain at rest if it is charge neutral. 

Having a macroion complex with given $N$, $q$ and $Q$, the interesting question
is not in which direction the effective force acts [since that question is
trivially answered by Eq.~(\ref{eqFtotal})] but rather whether the forces
become so large that the system disintegrates. (For a system with only gravity,
this problem is referred to as stability under tidal forces.)  This was noted
earlier when the dependence of the mobility of the macroion complex on an
applied field was analysed~\cite{lozada:99a,tanaka:02a}. While it was shown for
a few examples that some microions are ``ripped off'' the macroion, no
systematic study of the stability criterion under an applied external field has
been done (to the author's knowledge). Here, we aim to fill this gap.

With an applied external field $\vec{\mathcal{E}}$, Eq.~(\ref{eqVpot}) has
to be extended to
\begin{equation}
        V = \frac{q}{4\pi\epsilon R} \left[ 
                q \sum_{i<j}^{1\ldots N}
                \frac{1}{\abs{\vec{r}_i^{\circ} - \vec{r}_j^{\circ}}}
                - q \vec{E} \cdot \sum_{i=1}^N \vec{r}_i^{\circ} 
                + Q N \right]
        \label{eqVpotFeld}
\end{equation}
with the reduced electric field
\begin{equation}
        \vec{E} = \frac{4\pi\epsilon R^2}{q} \vec{\mathcal{E}} \;.
\end{equation}

The concept of \emph{globally stability} introduced in Sec.~\ref{secGlobal}
cannot be applied in the presence of an external field since the potential
energy is not bounded from below~\footnote{Imagine the bare macroion moving
against the direction of the field to infinity, and the microions  in the
direction of the field.}, and \emph{local} stability of
the complex is the relevant concept. Again, we move particle $k$ by a distance
$\Delta \ll R$ away from the
macroion.
To first order
the
potential energy $V'$ of the new state then becomes
\begin{subequations}
\label{eqVperturbedFeld}
\begin{align}
        V'&= 
                V-\frac{q}{4\pi\epsilon R^2}\left[
                q \sum_{i\ne k}^{1\ldots N} \frac{1}{2 
                | \vec{r}_i^{\circ} -  \vec{r}_k^{\circ}|} + Q +
                q\vec{E} \cdot \vec{r}_k^{\circ} \right]
                \Delta \label{eqVperturbedFelda} \\
         &\equiv V - \frac{1}{R} h_k \Delta \;.
\end{align}
\end{subequations}
The third term in the bracket of Eq.~(\ref{eqVperturbedFelda}) is the
difference
to Eq.~(\ref{eqVpotStrich}). It
describes the interaction with the
external field and depends on the angle between the position of the particle
and the external field.

As in Sec.~\ref{secLocal} the macroion complex is locally stable if and only if
$h_k<0$ for all $k$. 
A closer inspection of Eq.~(\ref{eqVperturbedFelda}) and comparison to
Eq.~(\ref{eqVpotFeld}) shows that the big bracket 
no longer is directly related to the energy of the $k$-th particle (as it was
in Sec.~\ref{secLocal}) as the sign in front of $\vec{E} \cdot
\vec{r}_k^{\circ}$ is inverted~\footnote{This sign is easily
understood by noting that the energies due to macroion-microion interaction as
well as due to microion-microion interaction become smaller in magnitude when
the $k$-th microion is moved away but the energy due to the external field
becomes larger in magnitude.}. 

\begin{figure}
\centering
\epsfig{file=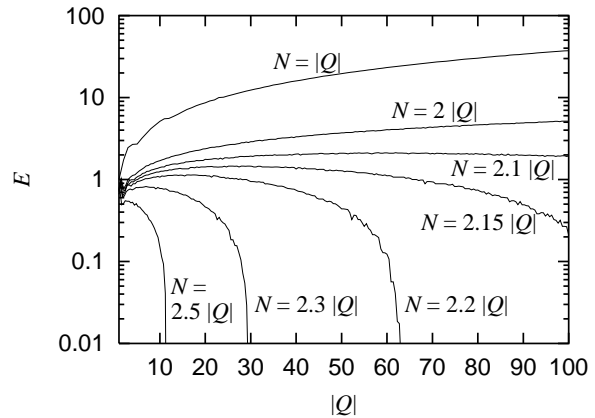,width=3.2in}
\caption{Electric field $E$ above which a state with $N$ microions bound to
the macroion is destroyed. (All labels ``$\abs{Q}$'' in the figure are to be
understood as $\abs{Q}/q$.)}
\label{qmin}
\end{figure}

We have been unable to find analytical expressions for the critical
external field at which the macroion complex becomes unstable and had to resort
to a numerical solution of Eq.~(\ref{eqVperturbedFeld}). The numerical
procedure, however, is basically identical to the one without an applied
external field, hence numerically very inexpensive. The result is depicted in
Fig.~\ref{qmin}. The roughness of the curves is not a sign of a numerical
problem but rather due to the physics of the problem. Depending on the
precise value for $N$, the geometrical arrangement is more or less symmetrical,
resulting in large changes in the dipole moment when $N$ is changed by only $1$.
Without an external field, this dipole moment is not relevant, and all
quantities are smooth functions of $N$. This is no longer the case now.

In Sec.~\ref{secLocal} we have shown that it is always possible to bind at
least $2\abs{Q}/q$ microions in a locally stable manner. Thus, for $N\le 2
\abs{Q}/q$ a finite electric field is necessary to break up the complex. For
larger $N$, however, the critical field may vanish, explaining the division of
the diagram into two separate regions by the line $N=2\abs{Q}/q$.

\begin{figure*}
\centering
\epsfig{file=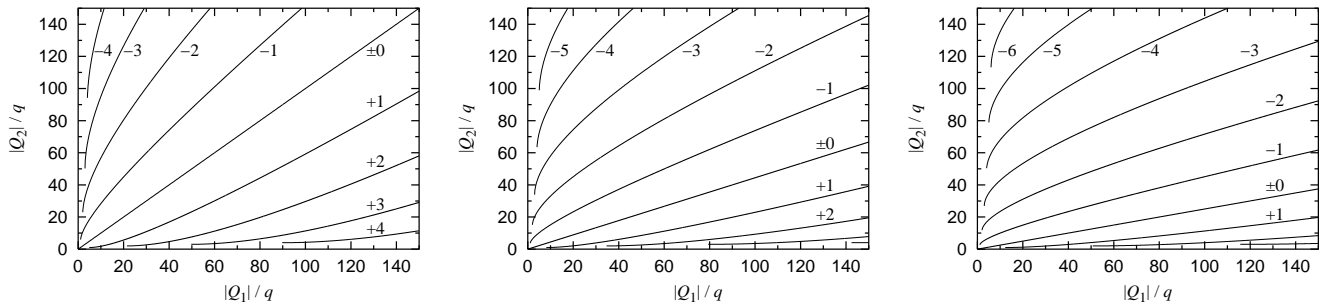,width=18.0cm}
\caption{Overcharging ($>0$) or undercharging ($<0$) 
of the first macroion (in units of $q$) as a function of the charges $Q_1$
and $Q_2$ of the two macroions. The radii of the two macroions are $R_1=R_2$
(left), $R_1=1.5 R_2$ (centre) and $R_1=2 R_2$ (left).
The lines are for the lowest-energy
configuration.}
\label{figAsymmetry}
\end{figure*}

\section{Charge distribution between two macroions}
\label{secAsymetry}

Next, we apply our approach to the
case of two well-separated macroions, with charges $Q_1$ and $Q_2$,
and radii $R_1$ and $R_2$,
respectively, together with $M\equiv \abs{Q_1+Q_2}/q$ microions so as to
achieve charge neutrality. According to Fig.~\ref{figLocal} and
Eq.~(\ref{eqUpper}), there are many different possibilities for distributing
those $M$ particles among the two macroions in a locally stable way. The naive
result is the one where 
each dressed macroion becomes charge neutral.
However, it is possible that $\delta$ microions will be transferred from
the first macroion to the second (if $\delta<0$, $\abs{\delta}$ ions are
transferred in the opposite direction). Such a state is called
``ionised''~\cite{messina:00a,messina:00b,messina:01b}.
Neglecting interactions among the two
macroions, since they are well separated, and applying Eq.~(\ref{eqPot3}) 
gives the potential energy $V(\delta)$ 
\begin{multline}
        V(\delta) = \frac{q^2}{4\pi\epsilon} \bigl[
                \frac{1}{R_1} f\left(\abs{Q_1}/q+\delta\right) +
                \frac{1}{R_2} f\left(\abs{Q_2}/q-\delta\right) \bigr] \\
                + \frac{q}{4\pi\epsilon} \bigl[ 
                        \left(\abs{Q_1}/q+\delta\right) \frac{Q_1}{R_1}
                        +  \left(\abs{Q_2}/q-\delta\right) \frac{Q_2}{R_2}
                \bigr] \;.
        \label{eqVzwei}
\end{multline}

We assume that the ions are at very low but finite temperature
such that the system can break out of a
locally stable state, and to find its lowest energy state~\footnote{Note that
this state is not globally stable even though it is the state of lowest energy.
In contrast to Sec.~\protect\ref{secGlobal}, where no energy barriers existed,
they do now.}.
This state is easily computed from Eq.~(\ref{eqVzwei}) and provides a relation
between $Q_1$, $Q_2$ and $\delta$ for the ground state. This result is most
easily displayed when $Q_2$ is expressed in terms of the other parameters,
\begin{multline}
        \abs{Q_2} = \delta + ( \delta + \abs{Q_1} ) \frac{R_2^2}{R_1^2}
                + \frac{4 \delta^2}{9 c^2} \left( 1 + \frac{R_2}{R_1} \right)^2
           \\
                - 4 \frac{\delta\sqrt{\abs{Q_1}+\delta}}{c}  
                        R_2 \frac{R_1+R_2}{R_1^2} \;.
        \label{eqQ2}
\end{multline}
This curve is depicted in Fig.~\ref{figAsymmetry} for different ratios of $R_1$
and $R_2$. For two identical
macroions we find that the lowest-energy state is the non-ionised one whereas
for $Q_1\ne Q_2$ but $R_1=R_2$ the macroion with higher charge attracts
more microions than
naively expected. This is in agreement with earlier results for $R_1=R_2$
where the
correlation effects in a three-dimensional layer were approximated by the
effects in a two-dimensional Wigner
crystal and confirmed by numerical
simulations~\cite{messina:00a,messina:00b,messina:01b}. 
For $R_1\ne R_2$ the macroions are uncharged only if 
\begin{equation}
        \frac{Q_1}{R_1^2} = \frac{Q_2}{R_2^2} \;.
        \label{eqUncharged}
\end{equation}
Other configurations than the lowest energy state can be excited thermally, and,
due to local stability, can persist for relatively long times.

In a strict mathematical sense, at finite temperature 
microions cannot be bound 
to a three-dimensional structure like a sphere. This is in contrast to 
a rod or a planar geometry~\cite{gelbart:00a}. If the temperature is not
too high, or equivalently, if the electrostatic coupling is strong enough
(i,\,e., large $\abs{Q_1}$, $\abs{Q_2}$ and $q$ as well as small $R$), the
methods presented in this paper can still be applied in an approximate fashion.
Microions stay very close to one macroion for most of the time, before they
hop (i.\,e., move within a time that is short compared to the time that they
remain effectively bound) to the other macroion. If we neglect the short hopping
phases, the probability for a given ionisation level $\delta$ is then given
by the Boltzmann factor, $P(\Delta) \propto \exp[-\beta V(\delta)]$,
and is easily evaluated numerically for arbitrary parameters.

\section{Conclusions}
\label{secConclusions}

To conclude, using general stability concepts we have shown that for a charged
macroion a stable overcharged state persists at zero temperature. The number of
globally stable microions is given by Eq.~(\ref{eqNstabile}), the number of
locally stable microions is given by Eq.~(\ref{eqUpper}). Physically, global
stability means that a random arrangement of microions will move to form a
layer around the macroion, whereas local stability means that a layer that
exists due to initial conditions will persist forever. An applied external
field, as used in electrophoresis, decreases the stability of the microion
cloud, as demonstrated in Fig.~\ref{qmin}.  We have also applied this approach to a
system of two macroions. In its ground state each of the two macroions acquires
a nonvanishing net charge, see Eq.~(\ref{eqQ2}), unless Eq.~(\ref{eqUncharged})
is fulfilled.

The value of our approach lies in the fact that it is exact, and by that
clarifies the effects caused by finite size, finite charge, correlations and
electrostatic interactions.

\acknowledgments

This work has been supported by the Academy of Finland Grant 
No.~54113 (M.\,K.), the Finnish Academy of Science and Letters (M.\,K.),
and by the Academy of Finland Centre for Excellence Program (2000-2005)
project no.~44897 (M.\,Patriarca).


\begin{thebibliography}{31}
\expandafter\ifx\csname natexlab\endcsname\relax\def\natexlab#1{#1}\fi
\expandafter\ifx\csname bibnamefont\endcsname\relax
  \def\bibnamefont#1{#1}\fi
\expandafter\ifx\csname bibfnamefont\endcsname\relax
  \def\bibfnamefont#1{#1}\fi
\expandafter\ifx\csname citenamefont\endcsname\relax
  \def\citenamefont#1{#1}\fi
\expandafter\ifx\csname url\endcsname\relax
  \def\url#1{\texttt{#1}}\fi
\expandafter\ifx\csname urlprefix\endcsname\relax\def\urlprefix{URL }\fi
\providecommand{\bibinfo}[2]{#2}
\providecommand{\eprint}[2][]{\url{#2}}

\bibitem[{\citenamefont{Moreira and Netz}(2000)}]{moreira:00a}
\bibinfo{author}{\bibfnamefont{A.~G.} \bibnamefont{Moreira}} \bibnamefont{and}
  \bibinfo{author}{\bibfnamefont{R.~R.} \bibnamefont{Netz}},
  \bibinfo{journal}{Europhys. Lett.} \textbf{\bibinfo{volume}{52}},
  \bibinfo{pages}{705} (\bibinfo{year}{2000}).

\bibitem[{\citenamefont{Manning}(1969{\natexlab{a}})}]{manning:69a}
\bibinfo{author}{\bibfnamefont{G.~S.} \bibnamefont{Manning}},
  \bibinfo{journal}{J. Chem. Phys.} \textbf{\bibinfo{volume}{51}},
  \bibinfo{pages}{924} (\bibinfo{year}{1969}{\natexlab{a}}).

\bibitem[{\citenamefont{Manning}(1969{\natexlab{b}})}]{manning:69b}
\bibinfo{author}{\bibfnamefont{G.~S.} \bibnamefont{Manning}},
  \bibinfo{journal}{J. Chem. Phys.} \textbf{\bibinfo{volume}{51}},
  \bibinfo{pages}{934} (\bibinfo{year}{1969}{\natexlab{b}}).

\bibitem[{\citenamefont{Oosawa}(1970)}]{oos:70}
\bibinfo{author}{\bibfnamefont{F.}~\bibnamefont{Oosawa}},
  \emph{\bibinfo{title}{Polyelectrolytes}} (\bibinfo{publisher}{Marcel Decker},
  \bibinfo{address}{New York}, \bibinfo{year}{1970}).

\bibitem[{\citenamefont{Shklovskii}(1999)}]{shklovskii:99a}
\bibinfo{author}{\bibfnamefont{B.~I.} \bibnamefont{Shklovskii}},
  \bibinfo{journal}{Phys. Rev. E} \textbf{\bibinfo{volume}{60}},
  \bibinfo{pages}{5802} (\bibinfo{year}{1999}).

\bibitem[{\citenamefont{Park et~al.}(1999)\citenamefont{Park, Bruinsma, and
  Gelbart}}]{park:99a}
\bibinfo{author}{\bibfnamefont{S.~Y.} \bibnamefont{Park}},
  \bibinfo{author}{\bibfnamefont{R.~F.} \bibnamefont{Bruinsma}},
  \bibnamefont{and} \bibinfo{author}{\bibfnamefont{W.~M.}
  \bibnamefont{Gelbart}}, \bibinfo{journal}{Europhysics Lett.}
  \textbf{\bibinfo{volume}{46}}, \bibinfo{pages}{454} (\bibinfo{year}{1999}).

\bibitem[{\citenamefont{Gelbart et~al.}(2000)\citenamefont{Gelbart, Bruinsma,
  Pincus, and Parsegian}}]{gelbart:00a}
\bibinfo{author}{\bibfnamefont{W.~M.} \bibnamefont{Gelbart}},
  \bibinfo{author}{\bibfnamefont{R.~F.} \bibnamefont{Bruinsma}},
  \bibinfo{author}{\bibfnamefont{P.~A.} \bibnamefont{Pincus}},
  \bibnamefont{and} \bibinfo{author}{\bibfnamefont{V.~A.}
  \bibnamefont{Parsegian}}, \bibinfo{journal}{Physics Today}
  \textbf{\bibinfo{volume}{53}}, \bibinfo{pages}{38} (\bibinfo{year}{2000}).

\bibitem[{\citenamefont{Grosberg et~al.}(2002)\citenamefont{Grosberg, Nguyen,
  and Shklovskii}}]{grosberg:02a}
\bibinfo{author}{\bibfnamefont{A.~Y.} \bibnamefont{Grosberg}},
  \bibinfo{author}{\bibfnamefont{T.~T.} \bibnamefont{Nguyen}},
  \bibnamefont{and} \bibinfo{author}{\bibfnamefont{B.~I.}
  \bibnamefont{Shklovskii}}, \bibinfo{journal}{Rev.~Mod.~Phys.}
  \textbf{\bibinfo{volume}{74}}, \bibinfo{pages}{329} (\bibinfo{year}{2002}).

\bibitem[{\citenamefont{Allahyarov et~al.}(1998)\citenamefont{Allahyarov,
  D'Amici, and L{\"o}wen}}]{allahyarov:98a}
\bibinfo{author}{\bibfnamefont{E.}~\bibnamefont{Allahyarov}},
  \bibinfo{author}{\bibfnamefont{I.}~\bibnamefont{D'Amici}}, \bibnamefont{and}
  \bibinfo{author}{\bibfnamefont{H.}~\bibnamefont{L{\"o}wen}},
  \bibinfo{journal}{Phys. Rev. Lett.} \textbf{\bibinfo{volume}{81}},
  \bibinfo{pages}{1334} (\bibinfo{year}{1998}).

\bibitem[{\citenamefont{Linse and Lobaskin}(1999)}]{linse:99a}
\bibinfo{author}{\bibfnamefont{P.}~\bibnamefont{Linse}} \bibnamefont{and}
  \bibinfo{author}{\bibfnamefont{V.}~\bibnamefont{Lobaskin}},
  \bibinfo{journal}{Phys. Rev. Lett.} \textbf{\bibinfo{volume}{83}},
  \bibinfo{pages}{4208} (\bibinfo{year}{1999}).

\bibitem[{\citenamefont{Mateescu et~al.}(1999)\citenamefont{Mateescu, Jeppesen,
  and Pincus}}]{mateescu:99a}
\bibinfo{author}{\bibfnamefont{M.}~\bibnamefont{Mateescu}},
  \bibinfo{author}{\bibfnamefont{C.}~\bibnamefont{Jeppesen}}, \bibnamefont{and}
  \bibinfo{author}{\bibfnamefont{P.}~\bibnamefont{Pincus}},
  \bibinfo{journal}{Europhys.~Lett.} \textbf{\bibinfo{volume}{46}},
  \bibinfo{pages}{493} (\bibinfo{year}{1999}).

\bibitem[{\citenamefont{Gurovitch and Sens}(1999)}]{gurovitch:99a}
\bibinfo{author}{\bibfnamefont{E.}~\bibnamefont{Gurovitch}} \bibnamefont{and}
  \bibinfo{author}{\bibfnamefont{P.}~\bibnamefont{Sens}},
  \bibinfo{journal}{Phys. Rev. Lett.} \textbf{\bibinfo{volume}{82}},
  \bibinfo{pages}{339} (\bibinfo{year}{1999}).

\bibitem[{\citenamefont{Lozada-Cassou et~al.}(1999)\citenamefont{Lozada-Cassou,
  Gonz{\'a}lez-Tovar, and Olivares}}]{lozada:99a}
\bibinfo{author}{\bibfnamefont{M.}~\bibnamefont{Lozada-Cassou}},
  \bibinfo{author}{\bibfnamefont{E.}~\bibnamefont{Gonz{\'a}lez-Tovar}},
  \bibnamefont{and} \bibinfo{author}{\bibfnamefont{W.}~\bibnamefont{Olivares}},
  \bibinfo{journal}{Phys. Rev. E} \textbf{\bibinfo{volume}{60}},
  \bibinfo{pages}{17} (\bibinfo{year}{1999}).

\bibitem[{\citenamefont{Nguyen et~al.}(2000{\natexlab{a}})\citenamefont{Nguyen,
  Grosberg, and Shklovskii}}]{nguyen:00a}
\bibinfo{author}{\bibfnamefont{T.~T.} \bibnamefont{Nguyen}},
  \bibinfo{author}{\bibfnamefont{A.~Y.} \bibnamefont{Grosberg}},
  \bibnamefont{and} \bibinfo{author}{\bibfnamefont{B.~I.}
  \bibnamefont{Shklovskii}}, \bibinfo{journal}{J. Chem. Phys.}
  \textbf{\bibinfo{volume}{113}}, \bibinfo{pages}{1110}
  (\bibinfo{year}{2000}{\natexlab{a}}).

\bibitem[{\citenamefont{Nguyen et~al.}(2000{\natexlab{b}})\citenamefont{Nguyen,
  Grosberg, and Shklovskii}}]{nguyen:00b}
\bibinfo{author}{\bibfnamefont{T.~T.} \bibnamefont{Nguyen}},
  \bibinfo{author}{\bibfnamefont{A.~Y.} \bibnamefont{Grosberg}},
  \bibnamefont{and} \bibinfo{author}{\bibfnamefont{B.~I.}
  \bibnamefont{Shklovskii}}, \bibinfo{journal}{Phys. Rev. Lett.}
  \textbf{\bibinfo{volume}{85}}, \bibinfo{pages}{1568}
  (\bibinfo{year}{2000}{\natexlab{b}}).

\bibitem[{\citenamefont{Messina
  et~al.}(2000{\natexlab{a}})\citenamefont{Messina, Holm, and
  Kremer}}]{messina:00a}
\bibinfo{author}{\bibfnamefont{R.}~\bibnamefont{Messina}},
  \bibinfo{author}{\bibfnamefont{C.}~\bibnamefont{Holm}}, \bibnamefont{and}
  \bibinfo{author}{\bibfnamefont{K.}~\bibnamefont{Kremer}},
  \bibinfo{journal}{Phys. Rev. Lett.} \textbf{\bibinfo{volume}{85}},
  \bibinfo{pages}{872} (\bibinfo{year}{2000}{\natexlab{a}}).

\bibitem[{\citenamefont{Messina
  et~al.}(2000{\natexlab{b}})\citenamefont{Messina, Holm, and
  Kremer}}]{messina:00b}
\bibinfo{author}{\bibfnamefont{R.}~\bibnamefont{Messina}},
  \bibinfo{author}{\bibfnamefont{C.}~\bibnamefont{Holm}}, \bibnamefont{and}
  \bibinfo{author}{\bibfnamefont{K.}~\bibnamefont{Kremer}},
  \bibinfo{journal}{Europhysics Lett.} \textbf{\bibinfo{volume}{51}},
  \bibinfo{pages}{461} (\bibinfo{year}{2000}{\natexlab{b}}).

\bibitem[{\citenamefont{Tanaka and Grosberg}(2001)}]{tanaka:01a}
\bibinfo{author}{\bibfnamefont{M.}~\bibnamefont{Tanaka}} \bibnamefont{and}
  \bibinfo{author}{\bibfnamefont{A.~Y.} \bibnamefont{Grosberg}},
  \bibinfo{journal}{J. Chem. Phys.} \textbf{\bibinfo{volume}{115}},
  \bibinfo{pages}{567} (\bibinfo{year}{2001}).

\bibitem[{\citenamefont{Messina et~al.}(2001)\citenamefont{Messina, Holm, and
  Kremer}}]{messina:01b}
\bibinfo{author}{\bibfnamefont{R.}~\bibnamefont{Messina}},
  \bibinfo{author}{\bibfnamefont{C.}~\bibnamefont{Holm}}, \bibnamefont{and}
  \bibinfo{author}{\bibfnamefont{K.}~\bibnamefont{Kremer}},
  \bibinfo{journal}{Phys. Rev. E} \textbf{\bibinfo{volume}{64}},
  \bibinfo{pages}{021405} (\bibinfo{year}{2001}).

\bibitem[{\citenamefont{Messina et~al.}(2002)\citenamefont{Messina, Holm, and
  Kremer}}]{messina:02a}
\bibinfo{author}{\bibfnamefont{R.}~\bibnamefont{Messina}},
  \bibinfo{author}{\bibfnamefont{C.}~\bibnamefont{Holm}}, \bibnamefont{and}
  \bibinfo{author}{\bibfnamefont{K.}~\bibnamefont{Kremer}},
  \bibinfo{journal}{Comp. Phys. Comm.} \textbf{\bibinfo{volume}{147}},
  \bibinfo{pages}{282} (\bibinfo{year}{2002}).

\bibitem[{\citenamefont{Terao and Nakayama}(2001)}]{terao:01a}
\bibinfo{author}{\bibfnamefont{T.}~\bibnamefont{Terao}} \bibnamefont{and}
  \bibinfo{author}{\bibfnamefont{T.}~\bibnamefont{Nakayama}},
  \bibinfo{journal}{Phys. Rev. E} \textbf{\bibinfo{volume}{63}},
  \bibinfo{pages}{041401} (\bibinfo{year}{2001}).

\bibitem[{\citenamefont{Neu}(1999)}]{neu:99a}
\bibinfo{author}{\bibfnamefont{J.~C.} \bibnamefont{Neu}},
  \bibinfo{journal}{Phys. Rev. Lett.} \textbf{\bibinfo{volume}{82}},
  \bibinfo{pages}{1072} (\bibinfo{year}{1999}).

\bibitem[{\citenamefont{Sader and Chan}(1999)}]{sader:99a}
\bibinfo{author}{\bibfnamefont{J.~E.} \bibnamefont{Sader}} \bibnamefont{and}
  \bibinfo{author}{\bibfnamefont{D.~Y.~C.} \bibnamefont{Chan}},
  \bibinfo{journal}{J. Colloid Interface Sci.} \textbf{\bibinfo{volume}{213}},
  \bibinfo{pages}{268} (\bibinfo{year}{1999}).

\bibitem[{\citenamefont{Trizac and Raimbault}(1999)}]{trizac:99a}
\bibinfo{author}{\bibfnamefont{E.}~\bibnamefont{Trizac}} \bibnamefont{and}
  \bibinfo{author}{\bibfnamefont{J.-L.} \bibnamefont{Raimbault}},
  \bibinfo{journal}{Phys. Rev. E} pp. \bibinfo{pages}{6530--6533}
  (\bibinfo{year}{1999}).

\bibitem[{\citenamefont{Evans and Wennerstr{\"o}m}(1999)}]{evans:99}
\bibinfo{author}{\bibfnamefont{D.~F.} \bibnamefont{Evans}} \bibnamefont{and}
  \bibinfo{author}{\bibfnamefont{H.}~\bibnamefont{Wennerstr{\"o}m}},
  \emph{\bibinfo{title}{The Colloidal Domain: Where Physics, Chemistry,
  Biology, and Technology Meet}} (\bibinfo{publisher}{Wiley},
  \bibinfo{address}{New York}, \bibinfo{year}{1999}), \bibinfo{edition}{2nd}
  ed.

\bibitem[{\citenamefont{Thomson}(1904)}]{thomson:04a}
\bibinfo{author}{\bibfnamefont{J.~J.} \bibnamefont{Thomson}},
  \bibinfo{journal}{Phil. Mag.} \textbf{\bibinfo{volume}{7}},
  \bibinfo{pages}{237} (\bibinfo{year}{1904}).

\bibitem[{\citenamefont{Erber and Hockney}(1991)}]{erber:91a}
\bibinfo{author}{\bibfnamefont{T.}~\bibnamefont{Erber}} \bibnamefont{and}
  \bibinfo{author}{\bibfnamefont{G.~M.} \bibnamefont{Hockney}},
  \bibinfo{journal}{J. Phys. A} \textbf{\bibinfo{volume}{24}},
  \bibinfo{pages}{L1369} (\bibinfo{year}{1991}).

\bibitem[{\citenamefont{Earnshaw}(1842)}]{earnshaw:42a}
\bibinfo{author}{\bibfnamefont{S.}~\bibnamefont{Earnshaw}},
  \bibinfo{journal}{Trans. Camb. Phil. Soc.} \textbf{\bibinfo{volume}{7}},
  \bibinfo{pages}{97} (\bibinfo{year}{1842}).

\bibitem[{\citenamefont{Bulatov}(1996)}]{bulatov:96a}
\bibinfo{author}{\bibfnamefont{V.}~\bibnamefont{Bulatov}}
  (\bibinfo{year}{1996}),
  \urlprefix\url{http://www.math.niu.edu/~rusin/known-math/96/repulsion}.

\bibitem[{\citenamefont{Drazin}(1992)}]{drazin:92}
\bibinfo{author}{\bibfnamefont{P.~G.} \bibnamefont{Drazin}},
  \emph{\bibinfo{title}{Nonlinear Systems}} (\bibinfo{publisher}{Cambridge
  University Press}, \bibinfo{year}{1992}).

\bibitem[{\citenamefont{Tanaka and Grosberg}(2002)}]{tanaka:02a}
\bibinfo{author}{\bibfnamefont{M.}~\bibnamefont{Tanaka}} \bibnamefont{and}
  \bibinfo{author}{\bibfnamefont{A.~Y.} \bibnamefont{Grosberg}}
  (\bibinfo{year}{2002}), \bibinfo{note}{cond-mat/0106561}.

\end{thebibliography}

\end{document}